\documentclass[12pt]{article}

\newcommand{\be}{\begin{equation}}
\newcommand{\ee}{\end{equation}}
\begin{document}
\title{ Possible way out of the Hawking paradox:  Erasing the information at the horizon}
\author{L. Herrera \thanks{Postal
address: Apartado 80793, Caracas 1080A, Venezuela.} \thanks{e-mail:
laherrera@cantv.net.ve} 
 \\
{\small Escuela de F\'{\i}sica, Facultad de Ciencias,}\\
{\small Universidad Central de Venezuela, Caracas, Venezuela.} \\
\\
}
\maketitle
\begin{abstract}
We show that small deviations from spherical symmetry, described by means of exact solutions to Einstein equations, provide a mechanism to ``bleach'' the information about the collapsing body as it falls through the aparent horizon, thereby resolving the information loss paradox. The resulting picture and its implication related to  the Landauer's principle in the presence of a gravitational field, is discussed.
\end{abstract}
\newpage

\section{Introduction}
The discovery by Hawking \cite{H1}, that quantum effects would cause a black hole to radiate and eventually to evaporate, together with the fact that such radiation is completely thermal (i.e.  it conveys no information) \cite{H2},  lead to two related contradictions, which form the well known information loss paradox.

On the one hand it follows that a pure  quantum state may evolve into a mixed state (the thermal radiation) in contradiction with the principle of unitary evolution. On the other hand, since the Hawking radiation is thermal, it appears that all information about the collapsing object is  lost forever.

So far this problem has been extensively discussed (see for example \cite{Presk}--\cite{Hebri} and references therein) but   no consensus has
been reached until now, concerning a satisfactory resolution of this quandary.

In this work we want to call the attention to a possible resolution of the loss information paradox, which is based on the fact that  close to the horizon   there is a bifurcation between any exact solution representing the external field of a perturbed sphere and the perturbed Schwarzschild spacetime (although strictly speaking the term ``horizon'' refers to the spherically symmetric case, we shall use it when considering the $r=2m$ surface, in the case of small deviations from sphericity),.

In the next section we shall elaborate on this issue, and  in section III we shall present the calculation of the area surface of the object for an exact solution to the Einstein equations representing a perturbed sphere. In section IV  we shall see that the evolution of the are surface, as the object contracts and approaches the horizon, will provide the  clue for the resolution of the information loss paradox that we propose.  Finally, some discussion is presented in the last section.
\section{The Israel theorem}
 It is well known \cite{1}, that the only static and asymptotically-flat vacuum space-time possessing a regular horizon is the Schwarzschild solution. For all
the others Weyl exterior solutions \cite{2}, the physical components of the Riemann tensor exhibit singularities at $r=2m$. This result is usually refererred to as the Israel
theorem.

On the other hand, we know that all physical systems are submitted to fluctuations and, of course,  this  also applies to self--gravitating systems.
Accordingly we have to assume that in the process of collapse (without angular momentum), the  spherical symmetry, is  permanently  submitted to such fluctuations.

Now, if the field produced by a self--gravitating system is not particularly intense (the boundary of the source is
much larger than the horizon) and fluctuations  off spherical symmetry are
sligth, then there is no problem in
representing the corresponding deviations from spherical symmetry (both inside and outside the source)
as a suitable perturbation of the spherically symmetric exact solution \cite{Letelier} . 

However, as the object becomes more
and more compact, such perturbative scheme will eventually fail near
the source. Indeed, as it is well known \cite{4}, as the boundary surface of
the source approaches the horizon,
any finite perturbation of the Schwarzschild spacetime, becomes
fundamentally different
from any exact solution, even if the latter is characterized by parameters
whose values are arbitrarily close to those corresponding to spherical
symmetry. In other words, for strong gravitational fields, no matter how
small the multipole moments (higher than monopole)
of the source are, there is a bifurcation between the perturbed Schwarzschild metric and all the other Weyl metrics. This of course is a consequence of the Israel theorem.

From the  above, a fundamental question arises: How should we describe the quasi--spherical space--time resulting from the fluctuations off Schwarzschild?
\begin{itemize}
 \item By means of a perturbed Schwarzschild metric,

 or 

 \item By means of an exact solution to Einstein equations, whose (radiatable) multipole moments are arbitrarily small, though non--vanishing.
 \end{itemize}

Our point of view is that the description of such deviations should be done from  an exact solution of Einstein equations (of the Weyl family, if we restrict ourselves to vacuum static
axially--symmetric solutions) continuously linked to the Schwarzschild metric through one of its parameters, instead of considering a perturbation of the Schwarzschild space--time.

It could be argued that  the black hole has no--hair theorem, according to which, in the process of
contraction all (radiatable) multipole moments are radiated away \cite{Price}, is  at variance  with that point of view.  However this is not so.

Indeed, the fact remains that  perturbations of spherical symmetry take place all along the evolution of the object. Thus, even if it is true that close to the horizon any of these
perturbations is radiated away, it is likewise true that this is a continuous process. In other words, as soon as a  ``hair'' is radiated away, a new  perturbation appears which will be later
radiated and so on. Therefore, since  ``hairs'' are radiated away at some {\it finite} time scale, then at that time scale there will be always a  fluctuation acting on the system (see \cite{h3} for a more detailed discussion on this point)

Thus,  if one wishes to describe the gravitational field of a
quasi-spherical source close to the horizon, one should use an
exact solution of Einstein equations,
rather than a perturbed Schwarzschild, no matter how small the
non-sphericity might be. If, for simplicity, one restricts oneself to the
family of axially symmetric non-rotating sources, then one has to choose
among the Weyl solutions.

The spacetime  to be considered here is the so-called
gamma
metric ($\gamma $-metric) \cite{9}, \cite{10}. This metric, which is
also
known as Zipoy-Vorhees metric \cite{11}, belongs to the family of Weyl's
solutions, and is continuously linked to the Schwarzschild space-time
through one of its parameters. The motivation for this choice stems from the fact  that  the $\gamma $-metric corresponds to a solution of the
Laplace equation (in cylindrical coordinates) with the same singularity
structure as the Schwarzschild solution (a line segment \cite{9}). In
this sense the $\gamma $-metric appears as the ``natural'' generalization of
Schwarzschild space-time to the axisymmetric case.  On the other hand it is worth noticing that physically meaningful sources for this metric have been  found \cite{gs}.

\section{The $\gamma$ metric}
In cylindrical
coordinates,
static axisymmetric solutions to Einstein equations are given by the
Weyl
metric \cite{2}
\begin{equation}
ds^2=e^{2\lambda }dt^2-e^{-2\lambda }\left[ e^{2\mu }\left( d\rho
^2+dz^2\right) +\rho ^2d\varphi ^2\right] ,  \label{1}
\end{equation}
with
\begin{equation}
\lambda _{,\rho \rho }+\rho ^{-1}\lambda _{,\rho }+\lambda _{,zz}=0
\label{2}
\end{equation}
and
\begin{equation}
\mu _{,\rho }=\rho \left(\lambda _{,\rho }^2-\lambda _{,z}^2\right)\qquad
\mu _{,z}=2\rho
\lambda _{,\rho }\lambda _{,z}.  \label{3}
\end{equation}
Observe that (\ref{2}) is just the Laplace
equation for $\lambda $ (in the Euclidean space).

The $\gamma $-metric is defined by \cite{9}
\begin{equation}
\lambda =\frac \gamma 2\ln \left[ \frac{R_1+R_2-2m}{R_1+R_2+2m}\right] ,
\label{4}
\end{equation}
\begin{equation}
e^{2\mu} =\left[ \frac{\left( R_1+R_2+2m\right) \left( R_1+R_2-2m\right)
}{%
4R_1R_2}\right] ^{\gamma ^2},  \label{5}
\end{equation}
where
\begin{equation}
R_1^2=\rho ^2+(z-m)^2\qquad R_2^2=\rho ^2+(z+m)^2.  \label{6}
\end{equation}
It is worth noticing that $\lambda ,$ as given by (\ref{4}), corresponds
to
the Newtonian potential of a line segment of mass density $\gamma/2 $ and
length $2m,$ symmetrically distributed along the $z$ axis. The
particular
case $\gamma =1,$ corresponds to the Schwarzschild metric.

It will be useful to work in Erez-Rosen coordinates, given by
\begin{equation}
\rho ^2=(r^2-2mr)\sin ^2\theta \qquad z=(r-m)\cos \theta ,  \label{7}
\end{equation}
which yields the line element as \cite{9}
\begin{equation}
ds^2=Fdt^2-F^{-1}\left\{ Gdr^2+Hd\theta ^2+\left( r^2-2mr\right) \sin
^2\theta d\varphi ^2\right\} ,  \label{8}
\end{equation}
where
\begin{eqnarray}
F &=&\left( 1-\frac{2m}r\right) ^\gamma ,  \label{8a} \\
&&  \nonumber \\
G &=&\left( \frac{r^2-2mr}{r^2-2mr+m^2\sin ^2\theta }\right) ^{\gamma
^2-1},
\label{8b}
\end{eqnarray}
and
\begin{equation}
H=\frac{\left( r^2-2mr\right) ^{\gamma ^2}}{\left( r^2-2mr+m^2\sin
^2\theta
\right) ^{\gamma ^2-1}}   \label{8c}
\end{equation}
Now, it is easy to check that $\gamma =1$ corresponds to the
Schwarzschild
metric.

The total mass of the source is \cite{9,10} $M=\gamma m,$ and the
quadrupole
moment is given by
\begin{equation}
Q=\frac \gamma 3m^3\left( 1-\gamma ^2\right) .  \label{9}
\end{equation}
So that $\gamma >1$ ($\gamma <1$) corresponds to an oblate (prolate)
spheroid.

Let us now calculate the area of the surface $r=const; t=const$ for the  line element (\ref{8}). We have

\begin{equation}
A=\int \sqrt{\mid g_{ij} \mid}d\phi d\theta,
\label{a1}
\end{equation}
where $\mid g_{ij}\mid$ is the determinant of the spatial metric. From (\ref{8}--\ref{8c}) and (\ref{a1}) it follows
\begin{equation}
A= 2\pi r^{2 \gamma} (r^2-2mr)^ {\frac{(\gamma-1)^2}{2}}\int _{0}^{\pi}\frac{\sin \theta}{(r^2-2mr+m^2 \sin^2 \theta)^{\frac{\gamma^2-1}{2}}} d\theta,
\label{a2}
\end{equation}
if $\gamma=1$ we obtain
\begin{equation}
A_{Sch.}= 4\pi r^{2},
 \label{a2}
\end{equation}
which is the well known result for the area surface in the Schwarzschild spacetime. However, if $\gamma \neq1$ the integral above is a hypergeometric function, producing
\begin{equation}
A= 4\pi r^{2 \gamma} (r^2-2mr)^ {\frac{(\gamma-1)^2}{2}}{(r-m)}^{(1-\gamma^2)}   {_2}F_{1}\left(\frac{1}{2},\frac{\gamma^2-1}{2}; \frac{3}{2}; \frac{m^2}{(r-m)^2}\right),
\label{a3}
\end{equation}
a result obtained in \cite{9} (except for a minor misprint in their equation (11)).
Now using the relationship \cite{denery} 
\begin{equation}
 {_2}F_{1}\left(a,b; c; 1\right)=\frac{\Gamma(c) \Gamma(c-b-a)}{\Gamma(c-b)\Gamma(c-a)},
\label{a4}
\end{equation}
where $\Gamma$ denotes the Gamma function, we see that  for $r=2m$, we have that  for values of $\gamma$ sufficiently close to $1$, i.e $\gamma=1+\epsilon$ with $\mid \epsilon \mid \ll 1$, $A\rightarrow 0$ as $r\rightarrow 2m$. Thus  as the object contracts  the area surface diminishes, vanishing for $r=2m$. This is true for both possible signs of $\epsilon$, even though the pattern  of evolution of $A$, is not the same in both cases. Indeed, in the case $\epsilon <0$ the ratio $\frac{A}{A_{Sch.}}$ monotonically decreases in the process of contraction, vanishing at the horizon. Instead, in the case $\epsilon >0$, the ratio  $\frac{A}{A_{Sch.}}$, initially increases until it reaches a maximum, and then start to decrease vanishing at the horizon.  These specific differences are irrelevant for our discussion, the important point being the vanishing of $A$ at the horizon.

Let us now relate this result with the information loss paradox.

\section{Bleaching of the information at the horizon}
 Let us first recall that according to the black--hole thermodynamics \cite{H4}--\cite{Bek4}, the entropy of the black--hole is related to the are surface of the horizon ($A_H$)  by  (in Plancks units $c=G=l_p=k_B=1$)
 \begin{equation}
 S=\frac{A_H}{4}.
 \label{area1}
\end{equation}

On the other hand this entropy is a measure of the information about the black hole interior \cite{Bek3}, implying that as the object with area $A$ contracts, the maximal information it can hold should satisfy the inequality \cite{Bek5}
\begin{equation}
I_{max} <A.
\label{area2}
\end{equation}

Let us now get back to the information loss paradox. 

One way to resolve the quandary would be to establish the  existence of correlations between the quanta emitted at different times, which would in principle carry all  of the information about the quantum state of the collapsing body. However, as stressed by Preskill \cite{Presk}, this  would necessarily imply  the bleaching of the information at the horizon, unless violation of  causality is allowed. 

If one assumes that the exterior spacetime is described by  the Schwarzschild metric (plus some small perturbations to take into account fluctuations of spherical symmetry), then one should not expect this bleaching to occur, since  for any freely falling observer the horizon is not a very special place. 

However, the situation is dramaticaly different if, in order to take into account the effects of fluctuations  on the spherical symmetry,  we describe the spacetime by means of an exact solution (no mater how close to the spherical symmetry), e.g. the $\gamma$-- metric with $\gamma=1+\epsilon$ with $\mid \epsilon \mid \ll 1$.  Now we can see that such a bleaching does in fact occur.

Indeed, in this later case, as the collapse proceeds, the area decreases according to (\ref{a3}),  vanishing at  the horizon, and so does the information held by the object, as it follows from  (\ref{area2}).

It is instructive to take a look on this issue from a different perspective. 

According to the Landauer's principle \cite{lan}, \cite{pie}, the erasure of  one bit of information stored in a system, requires  the dissipation into the environment of a minimal ammount of energy, whose lower bound is given by (in Planck units)
\begin{equation}
\bigtriangleup E=T \ln2,
\label{lan1}
\end{equation}
where $T$ denotes the proper temperature.
However if the system is located in a (weak) static gravitational field, the Landauer's principle  takes the form \cite{pla}
\begin{equation}
\bigtriangleup E=T(1+\phi) \ln2. 
\label{lan2}
\end{equation}
where $\phi$ denotes the (negative) gravitational potential, and  $T(1+\phi)$ (the Tolman's temperature) is the quantity  which is constant  in thermodynamic equilibrium \cite{Tol}

In the case of  a field of arbitrary strength,  Tolman's temperature becomes $T\sqrt{g_{tt}}$, accordingly   (\ref{lan2})  generalizes to
\begin{equation}
\bigtriangleup E=T \sqrt{g_{tt}}\ln2,
\label{lan3}
\end{equation}
implying that at the horizon (the $r=2m$ surface), either the proper temperature becomes singular or the erasure of information can be done without any dissipation of energy. If we exclude the  former possibility on physical  grounds, then we are left with the fact that $\bigtriangleup E$ should vanish at the horizon.

Now, as it follows from the information theory \cite{pie}, this situation (erasure without dissipation) corresponds to the case  where all  bits are already in one state only. Which is exactly the situation that emerges from the assumption that the radiation state is nearly pure, which in turn implies the bleaching of information at the horizon (see the discussion in pages 4 and 5 in \cite{Presk}). Thus the (modified) Landauer's principle seems to support  our conclusion about the bleaching of the information at the horizon.
\section{Conclusions}
We have seen that if the gravitational  field of the collapsing body,  including  small deviations from spherical symmetry,  is described by means of  an exact solution to Einstein field equations (instead of perturbations of Schwarzshild spacetime) then it  appears that  information about the collapsing body is stripped away, as the body falls through the horizon,   thereby resolving the information loss paradox.

We have also shown that  this picture is fully consistent with the Landauer's principle  in the presence of a gravitational field. Indeed, the fact that the  required energy to be dissipated,  in order to erase one bit of information at $r=2m$ vanishes, implies that when the object reach that point, is in a single state.

We are well aware of the fact that describing departures from sphericity  due to fluctuations, by means of exact solutions to Einstein equations instead of perturbations off Schwarzschild questions the very nature of black holes and by the same reason its thermodynamics.  However the important point  here is that  whatever the end state of a collapsing body  (whose surface boundary approaches $r=2m$) is, the emerging system will be deprived of any information. This is inferred by the Landauer's principle  and/or  assuming  the inequality  (\ref{area2}) to be valid.

 In this work we have restricted ourselves to a specific solution (the $\gamma$--metric) for reasons exposed at the Introduction. However, since there are a large number of 
different Weyl solutions which could be used to describe perturbations of spherical symmetry,  then
the  question arises:  which one among the Weyl solutions is
better entitled to describe small deviations from spherical symmetry?
It should be obvious that the question above has not a unique answer (there is an infinite number of ways of being non--spherical, so to speak). The important point to retain is that, for any Weyl solution (different from Schwarzschild), the horizon {\it is} a very special place. Therefore processes such as the bleaching of information discussed here, could be expected  in other Weyl spacetimes.  At any rate,  the choice of any specific Weyl metric  has to be reasoned.

Finally it is  worth noticing that we have referred exclusively to non--rotating sources. However we know that  a result similar to Israel theorem exists for
stationary solutions with respect to the Kerr metric \cite{carter}. Accordingly, it should
be expected that  a mechanism of bleaching of information,   related to fluctuations off Kerr (described by means of exact stationary solutions to Einstein
equations) would  exist, solving the information loss paradox in the general case. To prove that, is of course out of the scope of this work.
\section{Acknowledgements}
I  acknowledge financial support from the CDCH at Universidad Central de Venezuela under grant PI 03-00-7096-2008.

\end{document}